\documentclass[11pt]{article}

\usepackage{lscape,tikz}
\usepackage{epsfig,cancel,ulem, amsthm,amssymb}
\usepackage{leftidx}
\definecolor{darkblue}{rgb}{0.2,0.2,0.71}
\usepackage{todonotes}

\usepackage{color,soul}
\usetikzlibrary{decorations.markings,arrows}

\usepackage{wrapfig}
\usepackage{color}
\usepackage{graphicx,framed}
\usepackage[colorlinks=true, pdfstartview=FitV, linkcolor=darkblue, citecolor=darkblue, urlcolor=darkblue]{hyperref}
\definecolor{shadecolor}{rgb}{0.95, 0.95, 0.86}
\definecolor{darkgreen}{rgb}{0.2, 0.5,  0}

\def\&{\vspace{-5pt}&}

\def\Tr{ {\rm Tr}}

\def \eqref#1{(\ref{#1})}
\def \& {&\hspace{-10pt}}

\newcommand{\bt}{\beta}

\newcommand{\br}{{\mathbb R}}

   \newcommand{\bneg}{\mathfrak{b}}

  \newcommand{\gneg}{\mathfrak{m}}
  \newcommand{\nneg}{\mathfrak{n}}
  
  \newcommand{\lan}{\mathfrak{l}}

\newcommand{\g}{\mathfrak{g}}

 \newcommand{\ew}{\mathrm{W}}
 \newcommand{\exx}{\mathrm{E}}
  \newcommand{\A}{\mathcal{A}}
    \newcommand{\V}{\mathcal{V}}
\newcommand{\bil}[2]{{\langle #1 | #2\rangle}}
\newtheorem{theorem}{Theorem}[section]
\newtheorem{example}[theorem]{Example}
\newtheorem{exercise}[theorem]{Exercise}
\newtheorem{conjecture}[theorem]{Conjecture}

\newtheorem{lemma}[theorem]{Lemma}
\newtheorem{remark}[theorem]{Remark}

\newtheorem{proposition}[theorem]{Proposition}
\newtheorem{corollary}[theorem]{Corollary}
\newtheorem{definition}[theorem]{Definition}

\def\V {\mathcal V}

\def\bt{\begin{theorem}}
\def\et{\end{theorem}}
\def\bc{\begin{corollary}}
\def\ec{\end{corollary}}
\def\bx{\begin{example}}
\def\ex{\end{example}}
\def\bxr{\begin{exercise}\small}
\def\exr{\end{exercise}}
\def\bl{\begin{lemma}}
\def\el{\end{lemma}}
\def\bd{\begin{definition}}
\def\ed{\end{definition}}
\def\bp{\begin{proposition}}
\def\ep{\end{proposition}}

\def\br{\begin{remark}}
\def\er{\end{remark}}

\def\be{\begin{equation}}
\def\ee{\end{equation}}

\def\&{\hspace{-15pt}&}
\def\bea{\begin{eqnarray}}
\def\eea{\end{eqnarray}}

\def\1{{\bf 1}}

\def\z{\zeta}

\newcommand{\h}{\mathfrak{h}}

\newcommand{\RK}{\mathfrak{\rho}}

\newcommand{\ad}{\mathrm{ad}}

\makeatletter
\@addtoreset{equation}{section}
\makeatother

\begin{document}
\title{On integrability of transverse Lie-Poisson structures at nilpotent elements }
\author{Yassir Dinar}
\date{}

\maketitle

\begin{abstract}
We construct  families of functions in involution  for  transverse Poisson structures at nilpotent elements of  Lie-Poisson structures on simple Lie algebras by using the argument shift method. Examples show that these families  contain   completely integrable systems that consist of polynomial functions.  We provide a uniform construction of these integrable systems for an infinite family of distinguished nilpotent elements of semisimple type.
\end{abstract}

\maketitle
\tableofcontents
\section{Introduction}

Let $M$ be a manifold of dimension $n$ and $B$  a Poisson structure on $M$. Fix a point  $x\in M$, let  $2\RK_x$ be the rank of $B$ at $x$. Then, using  Weinstein splitting theorem, we fix a neighborhood $U$  centered at $x$  with coordinates $(q_1,\ldots,q_{\RK_x},p_1,\ldots,p_{\RK_x}, z_1,\ldots,z_{n-2\RK_x})$ such that on $U$
\be \label{Darboux} B=\sum_{i=1}^{\RK_x} {\partial\over \partial q_i}\wedge {\partial\over \partial p_i}+  {1\over 2}\sum_{i,j=1}^{n-2\RK_x} \phi_{kl}(z) {\partial\over \partial z_i}\wedge {\partial\over \partial z_j}\ee
where $\phi_{kl}(z)$ are smooth functions that depend only on $z_1,\ldots,z_{n-2\RK_x}$ and vanish at $x$ (\cite{AMV},\cite{LPV}). Note that the intersection of the symplectic leaf containing  $x$ with $U$ is defined by the vanishing of  $z_1,\ldots,z_{n-2\RK_x}$. The submanifold $N\subset U$ defined by
\[ N=\{x'\in U: q_i(x')=0=p_i(x'),~i=1,\ldots,{\RK_x}\}\]
is transverse to the symplectic leaf of $x$ at $x$. Then $(z_1,\ldots,z_{n-2\RK_x})$ are well defined coordinates on $N$ and $N$ inherits what is called transverse Poisson structure \cite{LPV}
\[ B^N=  {1\over 2}\sum_{i,j=1}^{n-2\RK} \phi_{kl}(z) {\partial\over \partial z_i}\wedge {\partial\over \partial z_j}.\]
 Assume that $N'$ is another submanifold transverse to the symplectic leaf of $x$ at $x$, then we can perform Dirac reduction (called Poisson-Dirac reduction in \cite{LPV}) of $B$ to get a Poisson structure $B^{N'}$ on  $N'$. Then there exists at $x$, a local Poisson diffeomorphism between $(N,B^{N})$ and  $(N', B^{N'})$  (\cite{LPV}, section 5.3).  In this article, we  consider the transverse Poisson structures of Lie-Poisson brackets at nilpotent elements.

The rank of $B$ is defined to be $2\RK=\max_{x\in M} 2\RK_x$. From the formula \eqref{Darboux}, we get  $2p=2\RK_x+\max_{t\in U}\, \textrm{rank}\, \phi_{kl}(t)$. In particular, the rank of the transverse Poisson structure $B^N$ at $x$ is $2\rho-2\rho_x$.

We call  a family  of functionally independent functions  $f_1,\ldots,f_{n-{\RK}}$  a (Liouville) completely integrable system if they are in  involution under $B$. Moreover, we say that the Poisson structure is polynomial if there exist coordinates where the entries of the matrix of $B$ are polynomial functions in the coordinates. In this case, a completely integrable system is called polynomial if it consists of polynomial functions. In this article, we provide examples of polynomial completely integrable systems for polynomial Poisson structures.

Let us assume that the rank of $B$ is constant on $U$, i.e. for all $y\in U$, $\RK_y=\RK$. Then we say that  $B$ is a regular Poisson structure on $U$ and the functions $\phi_{ij}(z)$ given in equation \eqref{Darboux} vanish on $U$.  In this case, the Poisson structure is also known as a constant Poisson structure (\cite{AMV},\cite{LPV}).  Then the coordinates given in \eqref{Darboux}   are called Darboux coordinates  and  the coordinate functions $q_1,\ldots, q_\rho,z_1,\ldots,z_{n-2\rho}$ form a polynomial  completely integrable system. In other words, integrability of constant or regular Poisson structure is obtained through Darboux coordinates.

Suppose  $B$ is a linear Poisson structure. Then $B$ is a Lie-Poisson structure on the dual space $\g^*$ of some Lie algebra $\g$ and it is not regular. Assume further that  $\g$ is a semisimple Lie algebra. Then $B$ can be defined on $\g$. In this case,   Miscenko and Fomenko construct a  polynomial completely integrable system for $B$ \cite{Mfomenko}. In their construction, they used a  compatible Poisson structure  $B_a$  with $B$  related to a regular semisimple element $a$ in $\g$. Then they apply what it is  known now as the argument shift method \cite{bolv} on the compatible Poisson structures to find a family of functions in involution. In the end, they proved that this family contains a sufficient number of independent function. We will review  this construction in section 2 below. For arbitrary linear Poisson structures, a conjecture known as  Miscenko-Fomenko conjecture states the existence of polynomial completely integrable systems for any linear Poisson structure. This conjecture was proved in \cite{SAD} using a different method than the argument shift method. More information about open problems concerning integrability of linear Poisson structures is given in   (\cite{bolv2}, section 5).

 The existence of polynomial completely integrable systems for constant and linear Poisson structures leads to the problem of finding examples of polynomial nonlinear   Poisson structures that admit polynomial completely integrable systems (this problem is also posed in a recent review paper by Bolsinov et al., see  (\cite{bolv2}, section 5.b).  In this paper, we give an infinite number of such examples. We consider transverse Poisson structures of Lie-Poisson structures on simple Lie algebras at nilpotent elements, and we use the argument shift method to construct  corresponding families of  polynomial functions in involution.

 The argument shift method for a bihamiltonian structure works as follows:  Assume on  $M$ there are compatible  Poisson structures $B_1$  and  $B_2$, i.e. $B_2$ and $B_1$ form a  bihamiltonian structure, and suppose that $B_1$ is in general position. Then, we  define the family of functions
 \be
\mathbf{F}[B_1,B_2]:=\cup_{\lambda\in\overline \mathbb{C}}\{F_\lambda: F_\lambda ~\mathrm{is~ a~ Casimir~ of} ~B_1+\lambda B_2\}.
\ee
This family commutes pairwise with respect to both Poisson brackets. Thus, we can find  a  completely integrable system by showing that  $\mathbf{F}[B_1,B_2]$ contains a sufficient  number of functionally independent functions. Bolsinov \cite{bolv}  proved that this is the case under certain condition on dimensions of singular sets of the Poisson pencils $B_1+\lambda B_2$, $\lambda\in \overline\mathbb{C}$. However,  methods used in this article are not using this result.

In order to formulate the main result in this article, let us assume that $M$ is a simple Lie algebra $\g$ and  $B$ is the Lie-Poisson structure on $\g$. Then the  symplectic leaves of $B$ coincide with the orbits of the adjoint group action. Let $L_1$ be a nilpotent element in $\g$. By Jacobson-Morozov theorem, there exist a nilpotent element $f$ and a semisimple element $h$ such that $A:=\{L_1,h,f\}\subseteq \g$ is  a $sl_2$-triple with relations
\begin{equation}  [h,L_1]= L_1,\quad [h,f]=-
f,\quad [L_1,f]= 2 h.
\end{equation}
We consider the  Slodowy slice $Q:=L_1+ \g^f$ where  $\g^f$ is  the subalgebra of centralizers of $f$ in $\g$. Then  $Q$ is a transverse subspace to the adjoint orbit of $L_1$, and it inherits the transverse Poisson structure $B^Q$ of  $B$ at $L_1$. It turns out that $B^Q$ is a polynomial  Poisson structure \cite{DamSab} (see also \cite{mypaper4} where an alternative proof is given by using the notion of bihamiltonian reduction and finite-dimensional version of Drinfeld-Sokolov reduction).
The rank of $B^Q$ in the case $L_1$ is regular or subregular nilpotent element is 0 and 2, respectively. Hence, integrability is trivial in those cases. However, the rank is greater than $2$ for other types of nilpotent elements.  For example, when $L_1$ is a nilpotent element of type  $D_{2m}(a_{m-1})$  and $\g$ is a Lie algebra of type $D_{2m}$, $m\geq 2$, the rank of $B^Q$ is $2m-2$ while $\dim Q$ is $4m-2$.  Thus, it is natural to ask about the existence of completely integrable systems for this large family of polynomial Poisson structures.

To apply the argument shift method, we consider a bihamiltonian structure on $\g$ formed by compatible Poisson structures  $B$ and $B_{K_1}$,  where the definition of  $B_{K_1}$ depends on a nilpotent element ${K_1}$ related to $L_1$ (see equation \eqref{lie-poiss}). Then one can perform bihamiltonian reduction to obtain a bihamiltonian structure  $B^Q$ and $B_{K_1}^Q$ on $Q$ \cite{mypaper4}. The collection $\mathbf{F}[B^Q,B_{K_1}^Q]$ of Casimirs of Poisson pencils $B_\lambda^Q:=B^Q+\lambda B_{K_1}^Q$ needed to perform the argument shift method is easy to describe. Let $P_1,...,P_r$ be a complete set of generators of the ring of invariant polynomials on $\g$.  Then the restrictions of the functions  $P_i$ to $Q+\lambda K_1$   are a complete set of independent Casimirs of the Poisson pencil $B_\lambda^Q$. Hence, to find a completely integrable system, it remains to investigate whether  $\mathbf{F}[B^Q,B_{K_1}^Q]$ contains a sufficient  number of independent functions. Examples show that this is always the case. However,  we provide a uniform proof only for some distinguished nilpotent elements of semisimple type. The proof includes the family of nilpotent elements of type $D_{2m}(a_{m-1})$ mentioned above.  It relies on the notion of opposite Cartan subalgebras, the weights of the adjoint action of $A$ at $\g$, and properties of the so-called quotient map.  Precisely, we prove the following

\bt \label{mainthm2}
Let $A=\{L_1,h,f\}$ be an $sl_2$-triple in a simple Lie algebra $\g$ of rank $r$ where $L_1$ is in one of the following  distinguished nilpotent orbits  of  semisimple type: $D_{2m}(a_{m-1})$, $B_{2m}(a_m)$,  $F_4(a_2)$,  $E_6(a_3)$,   $E_8(a_2)$ and $E_8(a_4)$ (if $L_1$ is of type $Z_r(a_i)$ then $\g$ is of type $Z_r$).

Let $Q:=L_1+\g^f$ be the Slodowy slice and consider the transverse Poisson structure $B^Q$ of the Lie-Poisson structure  on  $\g$.  Let $P_1,...,P_r$ be a complete set of homogeneous generators of the invariant ring under the adjoint group action. Assume  $K_1$ is an eigenvector of $\ad_h$ of the minimal eigenvalue such that $L_1+K_1$ is regular semisimple. Consider  the family of functions $\overline P_i^j$ on $Q$ defined by  the expansion
\be
P_i(x+\lambda K_1):=\sum_{j=0}^{\beta_i}\lambda^{j} \overline P_i^j(x),~\forall x\in Q,~i=1,...,r
\ee
Then the set of all   functions ${\overline P_i^j}$ are   independent and   form a polynomial  completely integrable system under $B^Q$.
\et

We organize the paper as follows. In section 2,  we fix some notations and review Miscenko-Fomenko construction of a polynomial completely integrable system for Lie-Poisson bracket on a simple Lie algebra. In section 3, we review the construction of a bihamiltonian structure on the Slodowy slice of an arbitrary nilpotent element  and we show how the argument shift method can be applied.  We give properties and identities related to distinguished nilpotent elements of semisimple type in section 4. In section 5, we prove  theorem \ref{mainthm2}.   In the last section, we give some remarks.

\section{Integrability of Lie-Poisson structure}

In this section, we fix notations and state some facts about simple Lie algebras. For completeness of this article,  we review the Miscenko-Fomenko construction of polynomial integrable systems for the Lie-Poisson bracket.

Let $\g$ be a complex simple Lie algebra of rank $r$ with the Lie bracket $[\cdot,\cdot]$, and denote the Killing form by $\bil . .$. Define the adjoint representation $\ad: \g \to \textrm{End}(\g)$ by $\ad_{g_1}(g_2):=[g_1,g_2]$. For $g\in \g$, then $\g^g$ denotes the centralizer of $g$ in $\g$, i.e. $\g^g:=\ker \ad_g$, and $\mathcal O_g$ denotes  the orbit of $g$ under the adjoint group action.  The element  $g$ is called nilpotent if $\ad_g$ is  nilpotent in $\textrm{End}(\g)$ and it is called regular if $\dim \g^{g}=r$. Any simple Lie algebra contains regular nilpotent elements. The set of all regular elements is open dense in $\g$.

Using  Chevalley theorem, we fix  a complete  system of homogeneous generators $P_{1},...,P_{r}$ of the  algebra of invariant polynomials under the adjoint group action.  We  assume throughout this article,  the  degree of $P_i$ equals $\nu_i+1$. The numbers $\nu_1,\nu_2,\ldots,\nu_r$ are known as the exponents of $\g$ and we suppose  they are given in a non decreasing  order, i.e.  $\nu_i\leq \nu_j$ if $i<j$.
 Consider  the adjoint quotient map
 \be \label{adjoint} \Psi:\g\to \mathbb{C}^r,~~~ \Psi(x)=(P_1(x),\dots,P_{r}(x)).\ee
From the work of Kostant in  \cite{kostpoly} the rank of  $\Psi$ at $x$ equals $r$ if and only if $x$ is regular element in $\g$.

 We define  the gradient $\nabla F: \g\to \g$ for a function $F$ on $\g $ by the formula
\be \label{gradient}
\frac{d}{dt_{t=0}}F(x+tq)=\bil {\nabla F(x)}{q},~\forall x, q\in \g.\ee
It is obvious that for any $x\in \g$ the rank of $\Psi$ at $x$  equals  the dimension of the vector space generated by $\nabla P_i(x)$.

Let $K_1$ be an element in $\g$. Then, we define the following  bihamiltonian structure  on $\g$ which consists of Lie-Poisson bracket  $\{.,.\}$ and the so called  frozen Lie-Poisson bracket $\{.,.\}_{K_1}$ \cite{MRbook}. We denote their Poisson structures  by $B$ and $B_{K_1}$, respectively. In formulas, for any two functions $F$ and $G$ and  $v\in T_x^* \g\cong\g$, we have
\begin{eqnarray}
 \label{lie-poiss}
\{F,G\}(x)&=&\bil {x}{[ \nabla G(x),\nabla F(x)]}; ~~B(x)(v)=[ x,v]\\\nonumber
\{F,G\}_{K_1}(x)&=&\bil {K_1}{[ \nabla G(x),\nabla F(x)]}; ~~B_{K_1}(x)(v)= [K_1,v].
\end{eqnarray}
We will use this bihamiltonian structure under different assumptions on  $K_1$. However, the following setup do not depend on the property of $K_1$ (\cite{bolv1}, section 2.2.4). Consider the Poisson structure $B_\lambda:=B+\lambda B_{K_1}$, $\lambda\in \mathbb{C}$. Then  $B_\lambda$ is isomorphic to $B$ by means of the linear transformation $x\to x+\lambda K_1$. Thus,  the rank of $B_\lambda$ equals $\dim \g-r$. Moreover, the tangent space of the symplectic leaf through  $x$ is spanned by the image of $\ad_{x+\lambda K_1}$. Hence,   the symplectic leaf of $B_\lambda$ through $x$  coincides with the orbit $ \mathcal{O}_{x+\lambda K_1}$.  Furthermore,  the polynomials  $ P_i(x+\lambda K_1)$ form   a complete set of  global independent  Casimir functions for $B_\lambda$, i.e.
\be \label{centerlamb}
[\nabla P_i(x+\lambda K_1), x+\lambda K_1]=0; ~~\forall i.
\ee
Following  the argument shift method, we consider the family of  functions
 \be
\mathbf{F}[B,B_{K_1}]:=\cup_{\lambda\in \mathbb{C}}\{F_\lambda: F_\lambda ~\mathrm{is~ a~ Casimir~ of} ~B+\lambda B_{K_1}\}.
\ee
We expand $P_i(x+\lambda K_1)$  in powers of $\lambda$
\be \label{expansion} P_i(x+\lambda K_1)=\sum_{j=0}^{\nu_i+1} \lambda^j P_i^j(x);~~\textrm{and} ~~\nabla P_i(x+\lambda K_1)=\sum_{j=0}^{\nu_i} \lambda^j\nabla P_i^j(x).
\ee
Then the functions $P_i^j(x)$ functionally generate $\mathbf{F}[B,B_{K_1}]$ and are in involution with respect to both Poisson brackets $\{.,.\}$ and $\{.,.\}_{K_1}$ \cite{Mfomenko}. Moreover,  $P_i^0(x)=P_i(x)$ and  $\nabla P_i^{\nu_i}(x)=\nabla P_i(K_1)$. Furthermore, using  \eqref{centerlamb}, we get the following  equations \be \label{recursive}
[\nabla P_i(x),x]=0,~[\nabla P_i^{\nu_i}(x),K_1]=0~\textrm{and}~[\nabla P_i^j(x),x]+[\nabla P_i^{j-1}(x),K_1]=0,~j=1,\ldots,\nu_i.
\ee
In particular,
\be \nabla P_i(x)\in \g^x ~\textrm{and}~ \nabla P_i^{\nu_i}(x)=\nabla P_i(K_1)\in \g^{K_1},~ \forall x\in \g,~i=1,\dots,r.\ee
As we mention in the introduction, to find a completely integrable system, it remains to prove that the set of functions $P_i^j(x)$ contains ${1\over 2}(\dim \g+ r)$ independent functions.

Let $L_1$ be a nilpotent element in $\g$. By Jacobson-Morozov theorem, there exist a nilpotent element $f$ and a semisimple element $h$ such that $A:=\{L_1,h,f\}\subseteq \g$ is  a $sl_2$-triple with relations
\begin{equation} \label{sl2:relation1} [h,L_1]= L_1,\quad [h,f]=-
f,\quad [L_1,f]= 2 h.
\end{equation}
Consider the Dynkin grading associated to $L_1$
\begin{equation}\label{grad1}
\g=\bigoplus_{i\in {1\over 2}\mathbb{Z}} \g_i;~~~~\g_i:=\{g\in\g: \ad_h g= i g\}
\end{equation}
It follows from representation theory of $sl_2$-algebras that  the eigenvalues of $ad_h$ are integers and half integers and
 $\ad_{L_1}: \g_i\to \g_{i+1}$ is surjective for $i>-1$ and injective  for $i<0$.

 Let us recall  Miscenko-Fomenko construction and the proof for the integrability on $B$.

\bt \cite{Mfomenko} \label{MP}
Assume $L_1$ is  regular and consider  the  expansion \eqref{expansion} with $K_1=h$. The  set of functions $\mathbf{T}=\{P_i^j(x)| i=1,\ldots,r,~ j=0,\ldots,\nu_i\}$ form  a polynomial completely integrable system for $B$.
\et
\begin{proof}
 We consider the bihamiltonian structure \eqref{lie-poiss} with $K_1=h$. Then  it is known that $h$ is a regular semisimple element and the eigenvalues of $\ad_h$ are all integers. In particular, $\h:=\g_0$ is a Cartan subalgebra. From the identities \eqref{recursive}, $\nabla P_i^{\nu_i}(L_1)=\nabla P_i(h)$ is  in $\h$ for every $i$. Since  $h$ is regular, properties of the adjoint quotient map implies that $\nabla P_i^{\nu_i}(L_1)$, $i=1,...,r$,  are linearly independent and form a basis for $\h$. The identities \eqref{recursive} also give $\ad_{L_1} \nabla P_i^j(L_1)=-\ad_h \nabla P_i^{j-1}(L_1)$. Since  $\ad_{L_1}:\g_i\to \g_{i+1}$ is surjective for $i\geq 0$,  the gradients  $\nabla P_i^{\nu_i-\alpha}(L_1)$ span the vector space $\g_\alpha$, $\alpha\geq 0$. Thus all gradients  of the functions in $\mathbf{T}$  span the space $\bneg=\oplus_{i\geq 0} \g_i$ which is a Borel subalgebra. Thus a lower bound for the number of functionally independent functions is $\dim \bneg$.    On the other hand, since $L_1$ is regular,  the restriction of the adjoint representation  to the $sl_2$-subalgebra $\A$ generated by $A$ decomposes $\g$ into $r$ irreducible $\A$-submodules $\V_i$ of dimensions   $2\nu_i+1$, $i=1,\ldots,r$. Each $\V_i$ has $\nu_i+1$ eigenvectors of $\ad_h$ of nonnegative eigenvalues. Then it follows from the definition of $\bneg$ that $\dim \bneg=\sum_{i=1}^r(\nu_i+1)$. Thus the cardinality of $T$ equals   the dimension of $\bneg$. Therefore,  elements of $\mathbf T$ are functionally  independent and  form a polynomial completely integrable system for $B$.
\end{proof}

\section{Argument shift method for transverse Poisson structure}
In this section, we review the construction of a bihamiltonian structure on Slodowy slice and we apply the argument shift method. We keep the notations introduced in the previous section.  We emphasize that the constructions and results in this article depend on the nilpotent orbit  $\mathcal{O}_{L_1}$ and not on the particular representative $L_1$ of  $\mathcal{O}_{L_1}$.

 We fix a good  grading $\g=\bigoplus_{i\in {1\over 2}\mathbb{Z}} \overline{\g}_i$ of $\g$ compatible with the
 $sl_2$-triple $A=\{L_1,h,f\}$, i.e. $L_1\in \overline \g_1$, $h\in \overline \g_0$ and
 $\ad_{L_1}: \overline \g_i\to \overline\g_{i+1}$ is surjective for $i>-1$ and injective  for $i<0$. See \cite{ELC} for the definition and classification of good gradings associated to nilpotent elements. Note that the Dynkin  grading defined in  \eqref{grad1} is a good grading.

 We fix  an isotropic subspace $\lan\subset \overline\g_{-{1\over 2}}$ under the symplectic bilinear form on $\overline \g_{-{1\over 2}}$ defined by $(x,y)\mapsto\bil {L_1}{[x,y]}$. Let $\lan'$ denote the symplectic complement of $\lan$ and
 introduce the following  nilpotent subalgebras
\begin{equation}
\gneg:=\lan\oplus \bigoplus_{i\leq -1} \overline\g_i;~~~~~~\nneg:=\lan'\oplus \bigoplus_{i\leq -1} \overline \g_{i}.
\end{equation}
Let  $\bneg $ denote the orthogonal complement of $\nneg$ under $\bil . .$.

 We consider the bihamiltonian structure \eqref{lie-poiss} and assume ${K_1}$  centralizes the subalgebra $\nneg$, i.e. $\nneg\subseteq \g^{K_1}
$. We can take ${K_1}$ to be of the minimal degree under the good grading.

Define Slodowy slice  $Q:=L_1+\g^f$. This affine subspace is transverse to the adjoint orbit of $L_1$. The space $\g^f$ is invariant under the action of $\ad_h$. Let $X_1,X_2,\ldots,X_n$ be a basis of $\g^f$ of eigenvectors under $\ad_h$. We introduce the coordinates $(z_1,\ldots,z_n)$ such that an element ${K_1}\in Q$ can be written in the form $L_1+\sum z_i X_i$.  Then we have the following theorem.

\bt \cite{mypaper4}
The space $Q$ inherits a  bihamiltonian structure $B^Q$, $B_{K_1}^Q$ from $B$, $B_{K_1}$, respectively. Moreover, $B^Q$ and  $B_{K_1}^Q$ are polynomial in the coordinates $(z_1,\ldots,z_n)$ and $B^Q$ is the transverse Poisson structure of $B$ at $L_1$. This bihamiltonian structure is independent of the choice of a good grading and isotropic subspace $\lan$. It can be obtained equivalently by using the bihamiltonian reduction with Poisson tensor procedure, Dirac reduction and a finite-dimensional version of the generalized  Drinfeld-Sokolov reduction.
\et

Details on bihamiltonian reduction can be found in \cite{CMP}.  Drinfeld-Sokolov reduction is initiated and applied for regular nilpotent elements in \cite{DS}. Generalizations to other nilpotent elements are obtained in   \cite{gDSh2},\cite{BalFeh1} (see also \cite{mypaper}). The relation between Drinfeld-Sokolov reduction and bihamiltonian reduction in the case of regular nilpotent elements is treated in  \cite{Pedroni2} where the Poisson tensor procedure is also initiated (also called the method of transverse subspace in \cite{LPM}). The relation between Drinfeld-Sokolov reduction and Dirac reduction is also mentioned in \cite{BalFeh1}. The fact that $B^Q$ is a polynomial Poisson bracket is proved in \cite{DamSab} (see also \cite{LPV}), and an alternative proof is given in \cite{mypaper4}.

 Throughout this article, we denote the restriction of $P_i$ to $Q$ by $\overline P_i^0$. We observe that for any $\lambda \in \mathbb{C}$ the pencil $B_\lambda^Q:=B^Q+\lambda B_{K_1}^Q$  is obtained by a Dirac reduction of $B_\lambda:=B+\lambda B_{K_1}$. Since the functions $P_i(x+\lambda K_1)$, $x\in \g$,  form   $r$ global independent  Casimirs for $B_\lambda$, it follows that the functions $\overline P_i^0(z+\lambda K_1)$, $z\in Q$, form  $r$ independent Casimirs for $B_\lambda^Q$. In particular, the rank of $B_\lambda^Q$ is less than or equal  $\dim Q-r$. Moreover, $B^Q$ is the transverse Poisson structure, hence rank $B^Q$ equals $\dim \g-r-\dim \mathcal O_{L_1}=\dim Q-r $. Thus the rank  $B_\lambda^Q$  equals $\dim Q-r$ for almost all values of $\lambda$. In particular, $B_\lambda^Q$ is zero in the case $L_1$ is a regular nilpotent element, and thus integrability of $B^Q$ does not make sense.

In the following, we would like to show how to realize  that the Casimirs of $B_\lambda^Q$ are the functions $\overline P_i^0(z+\lambda K_1)$ when we apply the Poisson tensor procedure. To this end, let us  summarize the construction of the Hamiltonian vector field of a function $F$ on $Q$ under $B_\lambda^Q$. Let $z\in Q$ and    identify $T_z^*Q$ with ${\g^{L_1}}$ using  the Killing  form $\bil . .$. Then we consider  $dF(z)$ as an element of $\g^{L_1}$. We extend $dF(z)$ to a covector $v_F\in T_z^*\g$ by requiring that
\begin{enumerate}
    \item The projection of $v_F$ to $ {\g^{L_1}}$ equals $dF(z)$, and
    \item $B_\lambda(v_F)\in  \g^f\simeq T_zQ$.
\end{enumerate}
It turns out that this extension $v_F$ is unique and can be calculated by solving recursive equations.   Then the value of the reduced Poisson structure  is given by the formula
\begin{equation}\label{lax}
B_\lambda^Q(dF)=B_\lambda(v_F)=[z+\lambda {K_1} ,v_F].
\end{equation}
Thus we get  a Lax representation of  any Hamiltonian vector field in $Q$ under $B_\lambda^Q$. The following proposition describes the Casimirs of $B_\lambda^Q$ using \eqref{lax}.
\bp
  The functions  $\overline P_i^0(z+\lambda {K_1})$, $z\in Q$, $i=1,\ldots,r$, form   a  set of independent global Casimirs for a generic $B_\lambda^Q$.
\ep
\begin{proof}
We follow a method given in (\cite{AMV}, page 68). Let $\phi:\g\to gl(\mathbb C^m)$ be any faithful matrix representation of $\g$ and  $F$  a function on $Q$. Consider the Lax representation \eqref{lax} under $\phi$.  Let  $z(t)$ denotes  the integral curve of $B_\lambda^Q(dF)$ and $\Tr S$ denotes the trace of $S$, $S\in \phi(\g)$.      Then
\begin{eqnarray*}
\{\Tr(\phi(z+\lambda K_1))^i,F\}_\lambda z(t)&=&{d\over dt} \Tr(\phi(z+\lambda K_1))^i= i  \Tr\big((\phi(z+\lambda K_1))^{i-1} {dz\over dt}\big)\\\nonumber
 & =& i \Tr\big( (\phi(z+\lambda K_1))^{i} \phi(v_F)-(\phi(z+\lambda K_1))^{i-1} \phi(v_F)
(\phi(z+\lambda K_1))\big)=0
 \end{eqnarray*}
Thus $\Tr(\phi(z+\lambda K_1))^i$ are Casimirs of  $B_\lambda^Q$. Since the functions $\Tr (\phi(g))^i$, $g\in \g $ generate the ring of invariant polynomials under the adjoint group action, we conclude that the restriction   of the invariant polynomials $P_i$ to $Q+\lambda K_1$ are Casimirs of $B_\lambda^Q$.

\end{proof}

Recall that the nilpotent element $L_1$ is called subregular if $\dim \g^{L_1}=r+2$ and any simple Lie algebra contains subregular nilpotent elements.

\bc
If $L_1$ is a subregular nilpotent element then $B^Q$ possesses a polynomial integrable system.
\ec
\begin{proof}
In case   $L_1$ is a subregular nilpotent element we have   $\dim Q=r+2$, rank $B^Q$ is $2$ and  $\overline P_1^0,\ldots, \overline P_r^0$ are Casimirs of $B^Q$.  Thus $(\overline P_1^0,\ldots, \overline P_r^0,F)$ form a polynomial completely integrable system  for any polynomial function $F$ on $Q$.
\end{proof}

Similar to the treatment in the last section, to apply  the  argument shift method for a general nilpotent element $L_1$, we consider the family of  functions
 \be
\mathbf{F}[B^Q,B_{K_1}^Q]:=\cup_{\lambda\in \mathbb{C}}\{F_\lambda: F_\lambda ~\mathrm{is~ a~ Casimir~ of} ~B^Q+\lambda B_{K_1}^Q\}.
\ee
 and we consider the expansion
 \be \label{arg:sht}
 P_i(z+\lambda {K_1})=\sum_{j=0}^{\gamma_i} \lambda^{j}\,  \overline P_i^j(z),~~z\in Q
 \ee
Then  the functions $\overline P_i^0(z)$ are Casimirs of $B^Q$,  $\overline P_i^{\gamma_i}$ are Casimirs of $B_{K_1}^Q$ and the functions   $\overline P_i^j$ are in involution with respect to both Poisson structures \cite{Mfomenko}. The  numbers $\gamma_i$ depends on $L_1$ and ${K_1}$. Thus, to find a polynomial completely integrable system it is enough to show that the functions  $\overline P_i^j$ contain $ {1\over 2}(\dim Q+r)$ functionally independent functions.  This is indeed the case for all examples we calculated. In other words we conjecture the following:

\begin{conjecture}
The non constant functions $\overline P_i^j$ of the expansion \eqref{arg:sht} are  independent over $Q$ and form a polynomial completely integrable system for $B^Q$.
\end{conjecture}

Here is an example to illustrate the procedure.

\bx It is known that the nilpotent orbits in the special Lie algebra $sl_m$ are in one to one correspondence with the partitions of $m$. We consider the Lie algebra  $sl_5$ and a nilpotent element $L_1$ corresponding to the partition $[3,2]$.  In contrary to the treatment in next sections, $L_1$ is not of semisimple type \cite{Elash}. Using standard procedure to obtain the  $sl_2$-triples \cite{COLMC}, we set
\[ L_1=\z_{1, 2} + \z_{2, 3} + \z_{4, 5}, ~h=\z_{1,1}- \z_{3,3}+{1\over 2}\z_{4,4}-{1\over 2}\z_{5,5},~f=2 \z_{2, 1} + 2 \z_{3, 2} + \z_{5, 4}\]
 where $\z_{i,j}$ denote the standard basis of $gl_5(\mathbb{C})$. Then points  of  Slodowy slice $Q$ will take the form
\be \left(
\begin{array}{ccccc}
 2 u_6 & 1 & 0 & 0 & 0 \\
 2 u_1-\frac{2 }{5}u_5 & 2 u_6 & 1 & u_8 & 0 \\
 u_4 & 2 u_1-\frac{2}{5} u_5 & 2 u_6 & u_7 & 2 u_8 \\
 2 u_3 & 0 & 0 & -3 u_6 & 1 \\
 u_2 & u_3 & 0 & u_1+\frac{4}{5}  u_5& -3 u_6
\end{array}
\right)
\ee
We take $K_1=\z_{3,1}$ which has the minimal degree under the Dynkin grading. Using the same notations given in \eqref{arg:sht}, we get the following  completely integrable system
\begin{eqnarray}
\overline P_1^0 & = & u_1+ 3 u_6^2,~~\overline P_3^1 = u_6,~~\overline P_4^1 = u_5-\frac{45 }{4}u_6^2+\frac{5 }{4}u_1, \\\nonumber
\overline P_2^0& = & u_4 -10 u_6^3+10 u_1 u_6-8 u_5 u_6+5 u_3 u_8,\\\nonumber
\overline P_3^0 & = & u_4
   u_6-10 u_6^4+2 u_5 u_6^2-\frac{2 }{3}u_1^2+\frac{8 }{75}u_5^2-\frac{2 }{5}u_1 u_5+\frac{1}{2}u_3 u_7+\frac{1}{2}u_2
   u_8, \\\nonumber
\overline P_4^0 &= & u_4 u_5 -90 u_6^5+100 u_1 u_6^3-10 u_5
   u_6^3-\frac{45}{4} u_4 u_6^2+25 u_3 u_8 u_6^2-10 u_1^2 u_6+\frac{8}{5} u_5^2 u_6 \\\nonumber
   & &-6 u_1 u_5 u_6-5
   u_3 u_7 u_6-5 u_2 u_8 u_6 +\frac{5 }{4}u_1 u_4-\frac{5 }{4}u_2 u_7-5 u_1 u_3 u_8-4 u_3 u_5
   u_8.
\end{eqnarray}
We make the following change of coordinates using the Casimirs of $B_{K_1}^Q$ which has the property that the inverse map is also a polynomial map
\be \label{nice coord} (w_1,w_2,\ldots,w_8):=(\overline P_1^0, \overline P_2^0,\overline P_3^1, \overline P_4^1,u_2,u_3,u_7,u_8).
\ee
Then  the nonzero terms of the transverse Poisson bracket $B^Q$ are
{\small \begin{eqnarray}
\{w_3,w_5\}& = &\frac{5}{6}w_5,~ \{w_3,w_6\}=\frac{5}{6}w_6,~ \{w_3,w_7\}=-\frac{5}{6}w_7,~ \{w_3,w_8\}=-\frac{5}{6} w_8, \\\nonumber
 \{w_4,w_5\}&= &-\frac{5}{6} \left(140 w_6 w_3^2+35 w_5 w_3+18 w_4 w_6\right),~ \{w_4,w_6\}=-\frac{25}{12}
   \left(w_5+4 w_3 w_6\right), \\\nonumber \{w_4,w_7\} & = & \frac{5}{6} \left(140 w_8 w_3^2+35 w_7 w_3+18 w_4 w_8\right),~ \{w_4,w_8\}=
   \frac{25}{12} \left(w_7+4 w_3 w_8\right) ,~ \{w_7,w_8\}=\frac{10 }{3}w_8^2, \\\nonumber
  \{w_5,w_6\} &=&
   -\frac{10 }{3} w_6^2,~\{w_5,w_8\} = \frac{5}{9} \left(1080
   w_3^3-110 w_1 w_3+64 w_4 w_3+3 w_2-21 w_6 w_8\right),  \\\nonumber
  \{w_6,w_7\}&=& \frac{5}{9} \left(1080 w_3^3-110 w_1 w_3+64 w_4 w_3+3 w_2-21 w_6 w_8\right),~ \{w_6,w_8\} =  \frac{1}{9}
   \left(-540 w_3^2+25 w_1-8 w_4\right),\\\nonumber \{w_5,w_7\}&= & \frac{2}{45} \left(59400 w_3^4-10250 w_1 w_3^2-2840 w_4 w_3^2+375 w_2
   w_3-3375 w_6 w_8 w_3-144 w_4^2\right)\\\nonumber & &+\frac{2}{45}(450 w_1 w_4+75 w_6 w_7+75 w_5 w_8).
\end{eqnarray}}
In particular the vector field
\be \chi_4=\sum_{i=5}^{8} \{w_4,w_i\}{\partial \over \partial w_i}\ee
is an integrable Hamiltonian vector field on $Q$.
\ex

\section{Distinguished nilpotent elements  of semisimple type}

In this section, we collect properties of distinguished nilpotent elements of semisimple type in simple Lie algebras and we derive important identities needed to prove our main results.

 Let us recall some definitions and notations from   \cite{COLMC}.  If $L_1$ is  regular then the orbit space $\mathcal{O}_{L_1}$ equals the set of all regular   nilpotent elements in  $\g$ (\cite{COLMC}, page 58). While if  $L_1$ is  subregular then the orbit space $\mathcal{O}_{L_1}$ equals the set of all  subregular nilpotent elements in  $\g$ (\cite{Tauv}, Proposition 34.5.7). The  nilpotent orbit  $\mathcal{O}_{L_1}$    is called {\bf distinguished}, and hence also $L_1$, if  $\mathcal{O}_{L_1}$ has no representative in a proper Levi subalgebra of $\g$. It turns out that $L_1$ is distinguished iff $\dim \g_0=\dim \g_1$. Also, when $L_1$ is distinguished, the eigenvalues of $\ad_h$ are all integers.  A regular nilpotent orbit is always distinguished.  Distinguished nilpotent orbits, along with other nilpotent orbits, are classified by using weighted Dynkin diagrams \cite{COLMC}. Distinguished nilpotent orbits are listed in the form $Z_r(a_i)$ where $Z_r$ is the type of  $\g$ and $i$ is the number of vertices of weight 0 in the corresponding weighted  Dynkin diagram.  If there is another orbit of the same number $i$ of 0's, then the notation $Z_r(b_i)$ is used. For example, regular nilpotent elements will be of type $Z_r(a_0)$, while distinguished subregular ones will be of type   $Z_r(a_1)$.

Let $\kappa$ denote the maximum eigenvalue of $\ad_h$. The nilpotent element  $L_1$ is said to be of {\bf semisimple type} \cite{Elash}, if there exists an element $g$ of the minimal eigenvalue $-\kappa$ under $\ad_h$ such that $L_1+g$ is semisimple. In this case $L_1+g$ is called  a cyclic element. When  $L_1$ is also distinguished, the element $L_1+g$ will be regular semisimple.   The list of distinguished nilpotent orbits of semisimple types  is given  in \cite{Elash} and \cite{DelFeher}. It consists of
\begin{enumerate}
    \item All regular nilpotent orbits  in simple Lie algebras (those of type $Z_r(a_0)$) and  subregular nilpotent orbits $F_4(a_1)$, $E_6(a_1)$, $E_7(a_1)$ and $E_8(a_1)$.
    \item Nilpotent orbits of type $B_{2m}(a_m)$, $D_{2m}(a_{m-1})$,  $F_4(a_2)$,  $F_4(a_3)$, $E_6(a_3)$,   $E_7(a_5)$, $E_8(a_2)$,$E_8(a_4)$, $E_8(a_6)$ and $E_8(a_7)$.
\end{enumerate}

We mention that nilpotent orbits in classical Lie algebras are classified by the partition of the dimension of the fundamental representations. In this article, $B_{2m}(a_m)$ corresponds to the partition $[2m+1,2m-1,1]$ when the Lie algebra  $\g$ is  $so_{4m+1}$ (type $B_{2m}$). While as usual in the literature,   $D_{2m}(a_{m-1})$  corresponds to the partition $[2m+1,2m-1]$ when   $\g$ is  $so_{4m}$ (type $D_{2m}$).

 For the remainder of this article, we assume $L_1$ is a distinguished nilpotent element of  type $Z_r(a_{r-s})$ (thus  $\g$ is of type $Z_r$), where $Z_r(a_{r-s})$ is one of the nilpotent orbits
 \begin{center}
 $B_{2m}(a_m)$, $D_{2m}(a_{m-1})$,  $F_4(a_2)$,   $E_6(a_3)$, $E_8(a_2)$, and $E_8(a_4)$.
 \end{center}
The number $s$ is introduced in this form in order to give  universal statements to all nilpotent orbits under consideration.

Since   $L_1$  is  of semisimple  type, we  fix an element $K_1\in \g_{-\kappa}$ such that the cyclic element  $Y_1:=L_1 +K_1$ is regular semisimple. In what follows, we give a general setup associated with cyclic elements following the work of Kostant for the case of cyclic elements associated with regular nilpotent elements \cite{kostBetti}.
 Let $\h':=\g^{Y_1}$ be the Cartan subalgebra  containing  $Y_1$. It is known as the opposite Cartan subalgebra. Then  the adjoint group element $w$ defined by
\begin{equation} \label{qcoxeter}
 w:=\exp {2\pi \mathbf{i}\over {\kappa+1}}\ad_h
\end{equation}
acts on $\h'$ as a representative of a regular conjugacy class $[w]$  in the Weyl group $W(\g)$ of $\g$ of order ${\kappa+1}$ (\cite{Elash} \cite{DelFeher}).

The  element $Y_1$ is an eigenvector of $w$ of eigenvalue $\epsilon$ where $\epsilon$ is the primitive $(\kappa+1)$th root of unity. We also define the multiset  $\exx(L_1)$ which consists of natural numbers  $\eta_i$, $i=1,\ldots,r$, such that  $\epsilon^{\eta_i}$'s are the  eigenvalues of the action of  $w$ on $\h'$. We call  $\exx(L_1)$ the exponents of the nilpotent element $L_1$. Our justification of the name exponents for $\exx(L_1)$ is  that, in this contest, the  exponents $\exx(\g)$ of the Lie algebra equal  the exponents of the regular nilpotent element \cite{kostBetti} (a nilpotent element of type $Z_r(a_0)$). The set $\exx(L_1)$ can be  found by combining the results in \cite{DelFeher},\cite{Elash} and \cite{springer}. In table 1, we list in the first 2 columns the elements of   $\exx(L_1)$ and in table 2 we list the exponents of $\g$. Note that we give the elements of  $\exx(L_1)$  in a particular order such that the following significant relation between $\exx(L_1)$ and $\exx(\g)$ is simple to state and its proof is given by examining table 1 and table 2.

\bl \label{relat-expon} The following formula gives a bijective map between  $\exx(\g)$ and $\exx(L_1)$
\[
 \nu_i=
 \left\{
  \begin{array}{ll}
    \eta_i, & i\leq s; \\
   \eta_i+\kappa+1, & i>s.
  \end{array}
\right.
\]
\el

\begin{table}
\begin{center}

\begin{tabular}{|c|c|c|c|}\hline
Orbit &  \multicolumn{3}{|c|}{$\ew(L_1)$}  \\\hline
$Z_r(a_{r-s})$ & $\eta_1,\eta_2,\ldots,\eta_s$ & $\eta_{s+1},\ldots,\eta_{r}$ & $\eta_{r+1},\ldots,\eta_{n}$ \\
\hline

$B_{2m}(a_{m})$ &  $1,3,\ldots,2m-1$& $1,3,\ldots,2m-1$ & $1,2,\cdots,2m-2; m-1,m$ \\
\hline
$D_{2m}(a_{m-1})$ &  $1,3,\ldots,2m-1;2m-1$& $1,3,\ldots,2m-3$ &$1,2,\cdots,2m-2$\\\hline

$F_4(a_2)$ & 1,5 & 1,5 &1,2,3,4\\\hline

$E_6(a_3)$& $1,4,5$ & 1,2,5 &$1,2,2,3,3,4$\\\hline
$E_8(a_2) $ & $1,7,11,13,17,19$ & 3,9 & $5,8,11,14$\\
$E_8(a_4) $ & $1,7,11,13$ & 2,4,8,14 &$3,5,5,7,7,9,9,11$ \\
\hline  & \multicolumn{2}{|c|}{$\exx(L_1)$}& \\\hline
\end{tabular}
\caption{ Weights and Exponents  of $L_1$}
\label{we}
\end{center}
\end{table}

\begin{table}
\begin{center}

\begin{tabular}{|c|c|}\hline
$\g$ & $\nu_1,\nu_2,\ldots,\nu_r$ \\
\hline
$B_{2m}$ &  $1,3,\ldots,4m-1$\\
\hline
$D_{2m}$ &  $1,3,\ldots,2m-1$ \\\
&$2m-1,2m-3,\ldots,4m-3$\\
\hline

$F_4$ & 1,5,7,11\\
\hline
$E_6$& $1,4,5,7,8,11$\\\hline
 $E_8$ & $1,7,11,13,17,19,23, 29$\\
\hline
\end{tabular}
\caption{  Exponents  of $\g$}
\label{ex}
\end{center}
\end{table}

Let  $Y_1,Y_2,\ldots, Y_{r}$ be a basis of  $\h'$ of eigenvectors of $w$ such that  $w(Y_i)=\epsilon^{\eta_i} Y_i$. Then the elements $Y_i$ will have the form
\be Y_i=L_{i}+K_{i}; ~~L_{i} \in \g_{\eta_i},~ K_{i}\in \g_{\eta_i-(\kappa+1)},~~i=1,\ldots,r.\ee
The commutators $[Y_i,Y_j]=0$ imply  that the set  $\{L_{1},\ldots,L_{r}\}$ generates a commutative subalgebra of $\g^{L_1}$. Hence upon considering the restriction of the adjoint representation of the $sl_2$-subalgebra $\A$ generated by $A=\{L_1,h,f\}$, the vectors  $L_{i}$ are maximal weight vectors of irreducible $\A$-submodules  of dimensions  $2\eta_i+1$. We observe   that the total number of irreducible $\A$-submodules  is $n:=3r-2s$. The numbers  $
\eta_{r+1},\ldots,\eta_{n}
$
are given in the third column of table 1.
Let us fix a  decomposition of $\g$ into irreducible $\A$-submodules, i.e.
\begin{equation}\label{decompo}
\g=\bigoplus_{j=1}^{n} \V_j
\end{equation}
where $\dim \V_j=2\eta_j+1$  and  $L_{i}$ is maximal  weight vector of $\V_i$ for $i=1,...,r$.  We found it more convenient to denote also  the maximum vectors of the remaining spaces $\V_j$ by $L_j$.  The numbers  $\eta_1,...,\eta_{n}$ are known in the literature as the weights of the nilpotent element $L_1$ and  will be  denoted $\ew(L_1)$. A procedure to obtain $\ew(L_1)$ for  nilpotent elements of type $D_{2m}(a_{m-1})$ and $B_{2m}(a_m)$ is given in \cite{mypaperw}.

Recall that  $\bil . .$ denotes the Killing form of $\g$. Let  us define the matrix of its  restriction to $\h'$
\be\label{Opp Cartan}
A_{ij}=\bil {Y_i}{Y_j}.
\ee
\bp\label{opp Cartan prop}
The matrix $A_{ij}$ is nondegenerate and  antidiagonal in the sense that  \[A_{ij}=0, ~{\rm if}~\eta_i+\eta_j\neq \kappa+1.\]
\ep
\begin{proof}
It follows from the properties of Cartan subalgebras that the matrix $A_{ij}$ is nondegenerate. We will use the fact that the matrix $\bil . .$ is a nondegenerate invariant bilinear form on $\h'$. Hence for any element $Y_i$ there exists an element  $Y_j$ such that $\bil {Y_i}{Y_j}\neq 0$. Using  the Weyl group element  $w$ defined in \eqref{qcoxeter}, we get the equality
\[\bil {Y_i}{Y_j}=\bil {w Y_i}{w Y_j}=\exp {2(\eta_i+\eta_j)\pi \mathbf{i} \over {\kappa+1}}\bil {Y_i}{Y_j}. \]
It forces  $\eta_i+\eta_j=\kappa+1$ in case $\bil {Y_i}{Y_j}\neq 0$.
\end{proof}

 Observe that  rank $r$ is even.  Let us set $m:=r/2$.

\bl \label{normlize1}
The elements  $Y_i,~i>1$  can be normalized such that the only nonzero values of $\bil . .$ on the basis $Y_i$ are given as follows
\be
\bil{Y_i}{Y_{m-i+1}}=\kappa+1, ~~~\bil{Y_{m+i}}{Y_{2m-i+1}}=\kappa+1~; ~~~i=1,...,m.
\ee
\el
\begin{proof}
Form the last lemma the elements $Y_1,...,Y_{r}$ can be grouped to subsets of 4 or 2 elements where the restriction of $\bil . .$ will be nondegenerate and has the  form
\be \left(
  \begin{array}{llll}
  0 & 0 & * & * \\
  0 & 0 & * & * \\\
  * & * & 0 & 0 \\\
  * & * & 0 & 0
  \end{array}
\right)\,\mathrm{or} ~\left(
  \begin{array}{ll}
 0 & * \\\
  * & 0
  \end{array}
\right),
 \ee
respectively. Using simple linear changes, they can be transformed to blocks of anti-diagonal matrices  without losing the fact that they are eigenvectors of the action of $w$ on $\h'$.
\end{proof}

 We assume that the elements of the basis $Y_i$ of $\h'$ are normalized and satisfy the hypothesis of the previous lemma. Then we get the following identities

 \bl \label{norm2}
\be
\bil{L_i}{K_{j}}=\eta_j(\delta_{i+j,m+1}+\delta_{i+j,3n+1})
\ee
\el
\begin{proof}
Recall that
 \be Y_i=L_{i}+K_{i}; ~~L_{i} \in \g_{\eta_i},~ K_{i}\in \g_{\eta_i-(\kappa+1)},~~i=1,...,r.\ee
Using the identity  $0=[Y_i,Y_j]=[L_i,K_j]+[K_i,L_j]$ with the invariant bilinear form yields
\be 0=\bil h {[L_i,K_j]+[K_i,L_j]}= (\kappa-\eta_j+1 )\bil {L_i}{K_j}+ \eta_j \bil {K_i} {L_j}. \ee
This equation with the normalization  $\bil {Y_i}{Y_j}=\bil {L_i}{K_j}+\bil {K_i}{K_j}=(\kappa+1)(\delta_{i+j,m+1}+\delta_{i+j,3m+1})$ yield the required identity.
\end{proof}

\section{Argument shift method and adjoint quotient map}

We keep the notations and assumptions given in the last section. We prove below that the argument shift method produces a polynomial completely integrable system for $B^Q$.  Observe that  the dimension of $Q$ is $n=3r-2s$ and $\kappa$ denotes  the maximum eigenvalue of $\ad_h$.

 Consider  the adjoint quotient map
 $ \Psi$ defined in \eqref{adjoint}. As we mentioned before,  Kostant in  \cite{kostpoly} proved that the rank of  $\Psi$ at $x$ equals $r$ if and only if $x$ is a regular element in $\g$.  Later, Slodowy showed that the rank of $\Psi$ is $r-1$ at subregular nilpotent elements \cite{sldwy2}. Finally, the rank of $\Psi$ at distinguished  nilpotent elements in $\g$ was computed by  Richarson   \cite{richard} except for nilpotent elements of type $E_8(a_2)$.  In this article, we proved  the rank  at  nilpotent elements of type  $E_8(a_2)$ is $6$ (see corollary \ref{rank}).

Under the normalization given in lemma \ref{norm2}, we  fix a basis $e_0,e_1,e_2,\ldots$ for $\g$ such that: \begin{enumerate}
    \item The first $n+r$ are given in the following order \be K_m,L_m,L_1,L_2,\ldots,L_{m-1};L_{m+1},\ldots,L_n,K_1,K_2,\ldots,K_{m-1}; K_{m+1},\ldots,K_{2m},\ee
    \item $\bil{e_i}{Y_1}\neq 0$ if and only if $i=0$ or $i=1$.
\end{enumerate}
Then we define  on $\g$ the linear coordinates
\be
x_i(g):=\bil {e_i}{g}, ~~i=0,1,2,\ldots
\ee
In what follows we will trace the dependence of the invariant polynomials on the coordinates $x_0$ and $x_1$. Note that the gradient $\nabla F$ of a function $F$ on $\g$ will be given by the formula $\nabla F=\sum {\partial F\over \partial x_i} e_i$. Moreover,  Since $Y_1$ is regular, the gradients  $\nabla P_i(Y_1)$ are linearly independent and in fact a basis of  $\h'$.  We use these remarks in the following lemma.

\bl \label{first norm}
  The matrix  with entries  ${\partial P_i \over \partial x_j}(Y_1)$, $i,j=1,...,r$, is non-degenerate. Moreover, $P_i$ have the following form
\be\label{respoly1} \label{respoly2}
P_i= R_i^1+R_i^2+R_i^3
\ee where
\be R_i^1=\sum_{\{a:a(\kappa+1)=\nu_i-\eta_r\}}\theta_{i,a}x_1^{a+1}x_0^{\nu_i-a}~~\mathrm{and}~~ R_i^2=\sum_{a=0}^{\nu_i-1}\sum_{j=2}^{2m}  c_{i,j,a}x_1^{a}x_0^{\nu_i-a}(x_j+x_{j+n-1})
\ee
and ${\partial R_i^3\over \partial x_k}(Y_1)=0, \forall k$. Here  $
c_{i,j,a}$ and $ \theta_{i,a}$ are complex numbers.
\el

\begin{proof}
Since $\nabla P_i(Y_1)\in \g^{Y_1}=\h'$ and  $\h'$ has basis $Y_i=L_i+K_i$, we get
\be
\nabla P_i(Y_1)=\sum_{j=1}^{2m} \overline C_{i,j} Y_j=\sum_{j=1}^{2m} \overline C_{i,j} ( L_j+ K_j)=C_{i,1}(e_0+e_1)+\sum_{j=2}^{2m}C_{i,j}(e_j+e_{n+j-1}).
\ee
Hence
\be \label{above}
C_{i,j}=\left\{
  \begin{array}{ll}
    {\partial P_i\over \partial x_j}(Y_1)={\partial P\over \partial x_{j+n-1}}(Y_1)
   ,    & 1<j\leq r\hbox{;} \\
     {\partial P_i\over \partial x_1}(Y_1)={\partial P_i\over \partial x_{0}}(Y_1), & j=1\hbox{;}
  \end{array}
\right.
\ee
and ${\partial P_i\over \partial x_\gamma}(
Y_1)=0$ for other indices  $\gamma $ not included in \eqref{above}. By  definition of the  coordinates and lemma \ref{norm2},   $x_j(Y_1)$ are all zero except $x_1(Y_1)=1$ and $x_0(Y_1)=\kappa$. Hence, for $1<j\leq 2m$, ${\partial P_i\over \partial x_j}(Y_1)\neq 0$ implies that   ${\partial P_i\over \partial x_j}$ must contain a term of the form
\be
 \sum_{a=0}^{\nu_i-1} c_{i,j,a} x_1^{a}x_0^{\nu_i-a-1}, ~~c_{i,j,a}\in \mathbb{C}
\ee
This gives the formula for $R_i^2$. Note that ${\partial R_i^2\over \partial x_1}(Y_1)=0$. Thus, for ${\partial P_i\over \partial x_1}(Y_1)$ to  be nonzero, it must contain   terms of the form $f_{i,a}=x_1^{a+1} x_0^{\nu_i-a}$. But then $a$ is constrained by the identity \be  {\partial f_{i,a}\over \partial x_1}(Y_1)=(a+1) \kappa^{\nu_i-a}={\partial f_{i,a}\over \partial x_{0}}(Y_1)=({\nu_i-a}) \kappa^{\nu_i-a-1}
\ee
This leads to  the formula for $R_i^1$. The condition on $R_i^3$ is a direct consequence from our analysis.
Finally,  the non-degeneracy condition  follows from the  fact that the vectors  $\nabla P_i(Y_1)$ are a basis for $\h'$.
\end{proof}

We observe that  $\g^{L_1}$ is orthogonal to $\g^f$ under $\bil . . $ \cite{wang}. Thus $(x_1,....,x_{n})$ form a coordinate system on $Q$. Also $Y_1\in Q$  and its coordinates  are given  by $(x_1,....,x_{n})=(1,0,0,\ldots,0)$. From lemma \ref{norm2}, $x_{0}(q)=\bil {K_m}{L_1}=\kappa$ is  constant for every $q\in Q$.

We set degree $ x_i$ equals $\xi_i+1$ where  $\xi_i$ denotes  the  eigenvalue of $e_i$ under $\ad_h$,   $i=1,\ldots,2m$. We recall  the following theorem  due to Slodowy.

\bt (\cite{sldwy2}, section 2.5) \label{slodowy}  The restriction $\overline P_i^0$ of   $P_i$  to $Q$ is quasi-homogeneous polynomial of  degree $\nu_i+1$.
\et

Then we get the following refinement of the last lemma. We introduce the numbers $\beta_i$ where $\beta_i=0$ for $i\leq s$ and $\beta_i=1$ for $i>s$.

\bp \label{norm-coord}
 The functions  $\overline P_i^0$  in the coordinates $(x_1,\ldots,x_n)$  take the form
 \be  \label{norm-coord eq}
\overline P_i^0(x_1,\ldots,x_n)=\sum_{\{j:\nu_i-\deg x_j=\beta_i(\kappa+1)\}} \widetilde{c}_{i,j}x_1^{\beta_i} x_j+\overline R_i^3(x), ~~\widetilde {c}_{i,j}\in \mathbb{C}.
\ee
 Here  $\frac{\partial \overline R_i^3}{\partial x_k}(Y_1)=0$  for $k=1,\ldots,n$. Moreover, the square  matrix  ${\partial \overline P_i^0\over \partial x_j}(Y_1)$; $i,j=1,\ldots,r$  is nondegenerate.
 \ep
 \begin{proof}
  We obtain the restriction $\overline P_i^0$ of $P_i$ to $Q$ by setting in \eqref{respoly1}, $x_0=\kappa$ and $x_k=0$ for $k>n$ . From the quasihomogeneity of $\overline P_i^0$ and  lemma \ref{first norm}
\be
\overline P_i^0 (x_1,\ldots,x_n)=\sum_{a=0}^{\nu_i-1}\sum_{\{j:a(\kappa+1)=\nu_i-\deg x_j\}} \widetilde{c}_{i,j,a}x_1^{a} x_j+\overline R_i^3 (x), ~~\widetilde {c}_{i,j,a}\in \mathbb{C}
\ee
where ${\overline R_i^3}$ is the  restriction of $R_i^3$ to $Q$. The condition     \eqref{respoly2} implies  ${\partial \overline R_i^3\over \partial x_k}(Y_1)=0$, $ k=1,\ldots,n$. Note that $\deg P_i-\deg x_j=a(\kappa+1)$. Using the relation  between the multisets $\exx (\g)$ and $\exx(L_1)$ observed in lemma \ref{relat-expon}, $a$ can only take  the values $\beta_i$ indicated in the statement. In other words the constants $c_{i,j,a}$ in   \eqref{respoly1}  are nonzero only if $a=\beta_i$. This gives the form \eqref{norm-coord eq}. For the nondegeneracy condition, note that the only possible value for the index $a$  in \eqref{respoly1} is  $a=\beta_i$ and so $x_0$ appear only with the power $\nu_i-\beta_i$. This implies that   ${\partial P_i\over \partial x_j}(Y_1)={\partial \overline P_i^0\over \partial x_j}(
Y_1)$, $i,j=1,\ldots,2m$. Thus  the determinant of the required matrix is nonzero. \end{proof}

 We apply the procedure given in section 3 for the nilpotent element $L_1$ by considering Dynkin grading and setting $u=K_1$. Then,  using equation \eqref{norm-coord eq} and $\bil{L_m}{K_1}=1$,  the expansion \eqref{arg:sht} will take the form
  \be
 P_i(x+\lambda K_1)=\sum_{\{j:\nu_i-\deg x_j=\beta_i(\kappa+1)\}} \widetilde{c}_{i,j}(x_1+\lambda)^{\beta_i} x_j+\overline R_i^3(x+\lambda K_1)=\overline P_i^0(x)+\lambda^{\beta_i} \overline P_i^{\beta_i}(x), ~~x\in Q.
 \ee
 In particular, we have the following coordinates.

 \bl \label{nice coordinates}
Consider the degrees of the coordinates $(x_1,\ldots,x_n)$ given in theorem \ref{slodowy}. Then there exists a quasihomogeneous change of coordinates  on $Q$ defined by
\be
t_i:=\left\{
  \begin{array}{ll}

    P_i^{\beta_i}=\partial_{x_1}^{\beta_i} \overline P_i^{0},~~i=1,\ldots,r;\\
 x_i,~~~~i=r+1,\ldots,n
 \end{array}
\right.
\ee
such that the degree of $t_i$ equals $\eta_i+1$. Moreover, these coordinates can be chosen such that  there exists one index $i_0$ such that $t_{i_0}$ is the only coordinate depending on $x_1$  and having  the form $t_{i_0}=x_1+ \textrm{nonlinear~ terms}$.
 \el

 \begin{proof}
Note that  $\partial_{x_1}^{\beta_i}\overline P_i^0$ will have the form
 \be \partial_{x_1}^{\beta_i}\overline P_i^0=\partial_{x_1}^{\beta_i}\overline R_i^3(
 x)+\sum_{\{j:\nu_i-\deg x_j=\beta_i(\kappa+1)\}} \overline{c}_{i,j} x_j,~~\overline{c}_{i,j}\in \mathbb{C},
 \ee
 where $\partial_{x_j}\partial_{x_1}^{\beta_i}\overline R_i^3(0)=0$. Then
 \be
 \partial_{x_j}\partial_{x_1}^{\beta_i}\overline P_i^0(0)=   {\partial \overline P_i^0\over \partial x_j}(Y_1), ~ i,j=1,\ldots,r.
 \ee
  Using the last proposition, we conclude that  the matrix   $\partial_{x_j}\partial_{x_1}^{\beta_i}\overline P_i^0$ is nondegenerate. Hence, $ \partial_{x_1}^{\beta_i}\overline P_i^0$ can replace the coordinates $x_i$ on $Q$ for $i=1,\ldots,r$. This shows that $(t_1,\ldots,t_n)$ defined above are  coordinates on $Q$. Quasihomogeneity and the degree of each $t_i$ follow from theorem \ref{slodowy} and lemma \ref{relat-expon}. This prove the first part.  The second part follows   from the structure of the  degrees  $\eta_i+1$ and the fact that $x_1$ is of maximum degree $\kappa+1$. If  there is another coordinate $x_{i_1}$ with maximum degree $\kappa+1$, then the statement follows by doing some linear change of coordinates. For example, for the  nilpotent element of type $E_8(a_4)$, we must take $i_0=r$.
 \end{proof}

 Note that $t_i= \overline P_i^0$ for $i=1,...,s$. Hence, in the coordinates developed in the last corollary,  the restriction $\overline \Psi$ of the quotient map $\Psi$ to $Q$ takes the form
 \be
 \overline \Psi(t_1,t_2,\ldots,t_n)=(t_1,...,t_{s},\overline P^0_{s+1},\ldots,\overline P^0_{r}).
 \ee

 \bc \label{rank}
 The rank of the quotient map $\Psi$ at $L_1$ equals  $s$.
 \ec
 \begin{proof}
From quasi-homogeneity,  the rank of $\Psi$ at $L_1$ is the same as the rank of $\Psi^Q$ at the origin  which equals  $s$.
 \end{proof}

Consider the set of functions $\mathbf T=\{\overline P_1^0,\ldots,\overline P_{s}^0,\overline P_{s+1}^1,\ldots,\overline P_{r}^1,\overline P_{s+1}^{0},\ldots,\overline P_{r}^0\}$ which result from applying the argument shift method to $B^Q$ and $B_{K_1}^Q$.    The set  $\mathbf T$ consists of   ${1\over 2}(\dim Q+ \textrm{rank } B^Q)=2r-s$ polynomial functions  in involution. We give below what remains to prove theorem \ref{mainthm2}.

\begin{proof}{[Theorem \ref{mainthm2}]}
We only need to prove that elements of $\mathbf T$ are functionally independent functions. For this task we use  properties of the restriction of the quotient map $\overline \Psi$. We consider what is called the momentum map $\Phi$ in the coordinates $(t_1,\ldots,t_n)$ developed above:
\begin{eqnarray}
\Phi(t_1,\ldots,t_{n})&:=&(\overline P_1^0,\ldots,\overline P_{s}^0,\overline P_{s+1}^1,\ldots,\overline P_{r}^1,\overline P_{s+1}^{0},\ldots,\overline P_{r}^0),\\\nonumber
&=&(t_1,\ldots,t_{r},\overline P_{s+1}^0,\ldots,\overline P_{r}^0).
\end{eqnarray}
Observe that the functions in $\mathbf T$ are independent if and only if the map $\Phi$ is regular at some points. The Jacobian matrices  $J\overline \Psi$ and $J\Phi$ of the maps $\overline \Psi$ and $\Phi$  have the forms
\begin{equation}
J\overline \Psi(t)=\left[\begin{array}{ccc}
I_s & 0 & 0\\
\phi_1 & \phi_2 & \phi_3
 \end{array}\right],~~~J\Phi(t)=\left[\begin{array}{ccc}
I_s & 0 & 0\\
0 & I_{r-s} & 0\\
\phi_1 & \phi_2 & \phi_3
\end{array}\right]
\end{equation}
where $I_m$ denotes the identity matrix of size $m$ and  $\phi_2$ is the square matrix of size $r-s$ with entries ${\partial \overline P^0_{s+i} \over \partial t_{s+j}}$; $1\leq i,j\leq r-s$. Then the entries of the matrices $\phi_1$ and $\phi_3$ are understood  from the definition of the Jacobian.  Thus, the  regularity of $\Phi$ is guaranteed  by showing that  the minor  matrix $\phi_3$  is of maximal rank at some points of $Q$. Let $V$ be the subvariety of $Q$ defined by  vanishing of the entries  of  $\phi_2$. By quasihomogeneity of the polynomials $\overline P_i^0$, entries of $\phi_2$ are not constant and hence $V$ is not empty. Since, the set of regular points of the quotient map is dense open subset in $\g$, and hence in $Q$, there is open subset  $U\subseteq V$ where the rank of $\overline \Psi$ is maximal. Thus $\phi_3$ is of maximal rank at the points of $U$. This ends the proof.
\end{proof}

\section{Remarks}

We observe that the proof of theorem \ref{mainthm2} depends on the fact that the argument shift method produces a set of functions with the property that each of them is either a Casimir of $B^Q$ or/and can be included as a part of coordinates on Slodowy slice. This property is not valid for  distinguished nilpotent orbits of semisimple type  $F_4(a_3)$,  $E_7(a_5)$,  $E_8(a_6)$ and $E_8(a_7)$.

In this article, we used the notion of opposite Cartan subalgebra and properties of the adjoint quotient map to prove integrability of transverse Poisson structure. However, examples show that integrability does not depend on these notions. Thus we believe that integrability of transverse Poisson structure at other types of nilpotent elements can be obtained using different methods than the ones given in this article.

We observe that even if the notion of Poisson reductions is well studied, there are no results about the reduction of completely integrable systems. For instance, it will be interesting to show that the restriction of the polynomial integrable system obtained by theorem \ref{MP} for $B$ leads to a completely integrable system of the transverse Poisson structure $B^Q$ on Slodowy slice $Q$.

Experts know that the linear terms of $B^Q$ define a  linear Poisson structure, which can be identified with the Lie-Poisson structure $B^{\g^f}$ of the Lie algebra $\g^f$.  Note that $\g^f$ is a nilpotent Lie algebra. In \cite{PY}, the existence of a completely integrable system for $\g^f$ using the argument shift method is proved for a large family of nilpotent elements. It will be interesting to find the relation between the method introduced in this article and \cite{PY}.

The quantization of $B^Q$ is studied in \cite{premet}, and it is known in the literature as a finite $W$-algebra \cite{wang}. Recently \cite{premet2}, the finite $W$-algebra is used to give a quantization of the completely integrable system that could be obtained for $B^{\g^f}$ using the argument shift method. It seems that the methods of \cite{premet2} can be used to obtain a quantization of the integrable system constructed in this article.

\noindent{\bf Acknowledgments.}

  A part of this work was done during the author visits to the Abdus Salam International Centre for Theoretical Physics (ICTP) and the International School for Advanced Studies (SISSA) through the years 2014-2017. This work was also funded by the internal grants of Sultan Qaboos University (IG/SCI/DOMS/15/04) and (IG/SCI/DOMS/19/08). The
author likes to thank anonymous reviewers  for critically reading the manuscript and suggesting substantial improvements.

\noindent Yassir Dinar

\noindent Sultan Qaboos University, Muscat, Oman

\noindent dinar@squ.edu.om

\end{document}